\documentclass[11pt]{article}
\usepackage{axodraw}
\usepackage{epsfig}
\usepackage{amsfonts}
\usepackage{amsmath}
\usepackage{bm,bbm}
\usepackage{cite}
\allowdisplaybreaks[1]

\input pix.sty

\newcommand{\nref}[1]{(\ref{#1})}  

\newcommand{\diag}{\mathop{\mbox{diag}}}
\newcommand{\f}{f}  
\newcommand{\sH}{\rmii{$H$}}
\newcommand{\I}{\rmii{$I$}}
\newcommand{\J}{\rmii{$J$}}

\newcommand{\T}{\rmii{$T$}}

\newcommand{\YB}{Y_\rmii{$B$}}

\renewcommand{\eq}{eq.~}
\renewcommand{\eqs}{eqs.~}
\renewcommand{\se}{sec.~}

\renewcommand{\fig}{fig.~}
\renewcommand{\figs}{figs.~}

\def\lsi{\raise0.3ex\hbox{$<$\kern-0.75em\raise-1.1ex\hbox{$\sim$}}}
\def\gsi{\raise0.3ex\hbox{$>$\kern-0.75em\raise-1.1ex\hbox{$\sim$}}}
\newcommand{\lsim}{\mathop{\lsi}}
\newcommand{\gsim}{\mathop{\gsi}}

\newcommand{\nF}{n_\rmii{F}}

\newcommand{\rmii}[1]{{\mbox{\tiny\rm{#1}}}}

\newcommand{\re}{\mathop{\mbox{Re}}}
\newcommand{\im}{\mathop{\mbox{Im}}}

\newcommand{\Tint}[1]{{\hbox{$\sum$}\!\!\!\!\!\!\!\int\,}_{\!\!\!\!\raise-0.9ex\hbox{$\scriptstyle{#1}$}}}
\newcommand{\Tinti}[1]{{{\Sigma}\!\!\!\!\raise0.3ex\hbox{$\int$}_\rmii{${#1}$}}}

\newcommand{\bi}{\begin{itemize}}
\newcommand{\ei}{\end{itemize}}
\newcommand{\hide}[1]{ }

\newcommand{\msl}[1]{\,\slash\!\!\!{#1}\,}
\def\TAsc(#1,#2)(#3,#4,#5)%
{\SetWidth{2.0}\CArc(#1,#2)(#3,#4,#5)\SetWidth{1.0}}
\def\Lwidth{3}

\def\TAgl(#1,#2)(#3,#4,#5){\SetWidth{2.0}\PhotonArc(#1,#2)(#3,#4,#5){\Lwidth}%
{6.283 #3 mul 360 div #4 #5 sub #4 #5 sub mul sqrt mul Tdensity mul}%
\SetWidth{1.0}}
\def\TLgl(#1,#2)(#3,#4){\SetWidth{2.0}\Photon(#1,#2)(#3,#4){\Lwidth}
{#1 #3 sub #1 #3 sub mul #2 #4 sub #2 #4 sub mul add sqrt Tdensity mul}%
\SetWidth{1.0}}

\renewcommand{\picc}[1]{\;\parbox[c]{60pt}{\begin{picture}(60,30)(0,0)
\SetWidth{1.0}\SetScale{1.3} #1 \end{picture}}\; }
\def\Lwidth{1.3}

\renewcommand{\pic}[1]{\;\parbox[c]{30pt}{\begin{picture}(30,30)(0,-3)
\SetWidth{1.0}\SetScale{0.8} #1 \end{picture}}\;}
\renewcommand{\picb}[1]{\;\parbox[c]{45pt}{\begin{picture}(45,30)(0,-3)
\SetWidth{1.0}\SetScale{0.8} #1 \end{picture}}\;}

\makeatletter \@addtoreset{equation}{section} \makeatother
\renewcommand{\theequation}{\arabic{section}.\arabic{equation}}
\makeatletter
\renewcommand\section{\@startsection {section}{1}{\z@}%
                                   {-5.5ex \@plus -1ex \@minus -.2ex}
                                   {2.3ex \@plus.2ex}%
                                   {\normalfont\large\bfseries}}
\renewcommand\subsection{\@startsection{subsection}{2}{\z@}%
                                     {-3.25ex\@plus -1ex \@minus -.2ex}%
                                     {1.5ex \@plus .2ex}%
                                     {\normalfont\normalsize\bfseries}}
\renewcommand\thesection {\@arabic\c@section}
\renewcommand\thesubsection   {\thesection.\@arabic\c@subsection}
\renewcommand{\@seccntformat}[1]{%
\csname the#1\endcsname.\hspace{1.0em}}
\makeatother

\begin{document}

\flushbottom

\begin{titlepage}

\begin{flushright}
May 2020
\end{flushright}
\begin{centering}
\vfill

{\Large{\bf
 Sterile neutrino dark matter via coinciding resonances
}} 

\vspace{0.8cm}

J.~Ghiglieri$^\rmi{a}$
 and 
M.~Laine$^\rmi{b}$
 
\vspace{0.8cm}

$^\rmi{a}$%
{\em
SUBATECH, Universit\'e de Nantes, IMT Atlantique, IN2P3/CNRS,\\
4 rue Alfred Kastler, La Chantrerie BP 20722, 44307 Nantes, France \\}

\vspace{0.3cm}

$^\rmi{b}$%
{\em
AEC, 
Institute for Theoretical Physics, 
University of Bern, \\ 
Sidlerstrasse 5, CH-3012 Bern, Switzerland \\}

\vspace*{0.8cm}

\mbox{\bf Abstract}
 
\end{centering}

\vspace*{0.3cm}
 
\noindent
It has been proposed that two resonances could coincide in the 
early universe at temperatures
$T \sim 0.2 ... 0.5$ GeV: one between two nearly degenerate GeV-scale
sterile neutrinos, producing a large lepton asymmetry through  
freeze-out and decays; another between medium-modified active neutrinos 
and keV-scale sterile neutrinos, converting the lepton asymmetry into 
dark matter. Making use of a framework which tracks three sterile neutrinos 
of both helicities as well as three separate lepton asymmetries, and 
scanning the parameter space of the GeV-scale species, we establish 
the degree of fine-tuning that is needed for realizing this scenario. 

\vfill

 
\end{titlepage}

\tableofcontents

%
\section{Introduction}

It has been a long-standing dream that a theoretical and experimental
understanding of the mechanism of neutrino mass generation could also 
help to solve outstanding cosmological mysteries, such as the existence
of baryon asymmetry and of dark matter abundance. 

In the present paper we approach this dream from a minimalistic and
phenomenologically oriented perspective. 
The existence of non-vanishing 
neutrino masses can be accounted for 
by incorporating right-handed neutrino fields
in the Standard Model Lagrangian,
\be
 \mathcal{L}^{ }_\rmi{new-SM}  \equiv  
 \mathcal{L}^{ }_\rmi{old-SM} 
 + \bar{\nu}^{ }_\rmii{R} i \msl{\partial} \nu^{ }_\rmii{R} 
 - 
 \bigl( 
 \bar{\nu}^{ }_\rmii{R}\, \tilde{\phi}_{ }^\dagger 
 h_{ }^{ }\, \ell^{ }_\rmii{L}
 + 
 \bar{\ell}^{ }_\rmii{L}\, h_{ }^\dagger  
 \tilde{\phi}\,  \nu^{ }_\rmii{R} 
 \bigr)
 - 
 \frac{1}{2}
 \bigl(
 \bar{\nu}_\rmii{R}^c M^{ }_{ } \nu^{ }_\rmii{R} 
 + 
 \bar{\nu}_\rmii{R}^{ }\, M^{\dagger}_{ } \nu^c_\rmii{R} 
 \bigr)
 \;. \la{L} 
\ee 
Here $\tilde{\phi}\, \equiv i \sigma^{ }_2 \phi^*$ is a conjugated
Higgs doublet, $\nu^c_\rmii{R}$ is a charge-conjugated right-handed
neutrino field with three generation indices, $\ell^{ }_\rmii{L}$ are 
the Standard Model lepton doublets, 
and $h$ is a $3\times 3$ matrix of neutrino Yukawa couplings. 

A singular value decomposition 
permits to write the Majorana mass matrix $M$ 
in \eq\nr{L} as 
$M^{ }_{ } = O \mathop{\mbox{diag}}(M^{ }_1, M^{ }_2, M^{ }_3) O^T$, 
where $M^{ }_\I \ge 0$ can be set in increasing order.
In the seesaw regime~\cite{ss1,ss2,ss3}, the $M^{ }_\I$ are close
to the physical masses of sterile neutrino mass eigenstates. 
In this paper we work under the (unconfirmed) assumption that 
$M^{ }_1 \approx 7$~keV is the mass of a long-lived dark matter 
candidate~\cite{observe1,observe2}, 
whereas $M^{ }_{2,3} \sim $~a few GeV represent short-lived
states that 
decay before primordial nucleosynthesis
(cf.,\ e.g.,\ refs.~\cite{Neff0,Neff} and references therein). 

The GeV-scale mass range for $M^{ }_{2,3}$ 
is motivated by many reasons.
First of all, 
it leads to a leptogenesis scenario~\cite{ars,as} which
could be testable through experiments at the intensity
frontier, such as SHiP~\cite{ship}.
Second, it leads to 
an interesting dark matter scenario~\cite{singlet,late},
in which the freeze-out and decays of the GeV-scale 
states play a key role. 

More precisely, it is known that if this model is used
for baryogenesis, then 
lepton asymmetry production can continue after sphaleron 
freeze-out~\cite{singlet,eijima}, 
and the final lepton asymmetries can be up to $\sim 10^3$ 
larger than the baryon asymmetry~\cite{degenerate}.  
However, as suggested by previous studies~\cite{shifuller,dmpheno,db}
and confirmed by a precise investigation~\cite{simultaneous}, 
such asymmetries are still $\sim 10^2$ too small to  
produce the correct dark matter abundance through
the Shi-Fuller mechanism~\cite{sf}. 
The suggestion of refs.~\cite{singlet,late} is that the missing 
orders of magnitude could originate from the low-temperature
freeze-out and decays of the GeV-scale sterile states. 

The purpose of the current paper is, on one hand, to verify that the
scenario laid out in 
ref.~\cite{late} can be realized and, on the other,
to quantify the degree of fine-tuning that it depends on. 
To this end, after reviewing the general
framework (cf.\ \se\ref{se:framework}), we define a set of 
``low-energy constants'' (cf.\ \se\ref{se:lecs}) which capture 
the essential aspects of the low-temperature solution. We then
scan the parameter space, establishing the part in which 
the low-energy constants obtain their 
desired values (cf.\ \se\ref{se:scans}). Finally, picking 
a point from the allowed domain, we show a successful 
scenario which yields (even a bit more than) the correct baryon
asymmetry and dark matter abundance (cf.\ \se\ref{se:example}).

%
\section{Review of framework}
\la{se:framework}

Our study is based on a set of evolution equations for ``slow variables''
that was established in ref.~\cite{simultaneous} and confirmed through 
a rigorous derivation in ref.~\cite{dbX}. In order to permit for 
computationally expensive numerical scans, we simplify here the 
equations by momentum averaging the density matrices of the GeV-scale sterile 
species. This implies that their momentum dependence  
is assumed to take the form
\be
 \rho^\pm_{\sH}(\vec{k}) 
 \;\simeq\; 
 \nF^{ }(\omega^{ }_{\sH}) 
 \, 
 \frac{Y^\pm_{\sH}}{Y^+_\rmi{eq}}
 \;, \quad
 Y^{+}_\rmi{eq} \; \equiv \; 
 \frac{ \int^{ }_{\vec{k}} \nF^{ }(\omega^{ }_{\sH}) }
      { s^{ }_\T }
 \;, \la{ansatz} 
\ee
where $\pm$ stands for helicity symmetry/antisymmetry; 
the subscript $H$ refers to the ``heavy'' sterile species; 
$\nF^{ }$ is the Fermi distribution; 
$\omega^{ }_\sH \equiv \sqrt{k^2 + M^2_\sH}$ with $k \equiv |\vec{k}|$
and $M^{ }_\sH \equiv (M^{ }_2 + M^{ }_3)/2$;
and $s^{ }_\T$ denotes the thermal entropy density 
of the Standard Model degrees of freedom.
Inserting \eq\nr{ansatz} into the evolution equations, 
the coefficients appearing in them get averaged in one 
of two possible ways, 
\be
 \langle ... \rangle^{ }_1 \; \equiv \; 
 \frac{ \int^{ }_{\vec{k}} (...) \nF^{ }(\omega^{ }_{\sH}) }
      { \int^{ }_{\vec{k}} \nF^{ }(\omega^{ }_{\sH}) }
 \;, \quad 
  \langle ... \rangle^{ }_2 
 \; \equiv \; 
 \frac{ \int_{\vec{k}}
  (...) \, \nF{}(\omega^{ }_{\sH}) 
 \, [ 1 - \nF{}(\omega^{ }_{\sH}) ] }{s^{ }_\T} 
 \;. \la{average} 
\ee
Averages of the second type, originating in connection with certain terms
proportional to lepton chemical potentials, become Boltzmann-suppressed
when $T \ll M^{ }_{\sH}$.

An essential aspect of the dynamics is that the GeV-scale species 
fall out of equilibrium, i.e.\  
$
 Y^+_\sH \neq \mathbbm{1}\,Y^+_\rmi{eq}
$
and
$
 Y^-_\sH \neq 0
$.
This also affects the expansion of the universe. Denoting 
\be
  x \; \equiv \; \ln\biggl( \frac{T^{ }_\rmi{max}}{T} \biggr)
 \;, 
 \quad
 \mathcal{J} \; \equiv \; \frac{{\rm d}x}{{\rm d}t}
 \;, 
\ee
and referring for details to sec.~3 of ref.~\cite{simultaneous}, 
the upshot is that rate coefficients get 
normalized through the Jacobian $\mathcal{J}$, and the evolution
is affected by entropy increase,  {\it viz.} 
\be
 \widehat{C} \;\equiv\; \frac{C}{\mathcal{J}}
 \;, 
 \quad
 Y'(x) \;\equiv\; 
 \bigl[ \partial^{ }_x + 
 \partial^{ }_x\ln(s^{ }_\T a^3) 
 \bigr]Y(x)
 \;, \quad 
 f'(x,k) \; \equiv \; 
 \frac{{\rm d}f(x,k(x))}{{\rm d}x}
 \;, 
\ee
where $a$ is the scale factor and 
$
 \partial^{ }_x\ln(s^{ }_\T a^3) = 3 \widehat{H} - 1/c_s^2
$, 
where $H$ is the Hubble rate and $c_s^2$ is the 
speed of sound squared. 
Furthermore 
$k(x) = k(x^{ }_0) a(x^{ }_0) / a(x)$ 
denotes a co-moving momentum. 
In thermal equilibrium, 
$\mathcal{J} = 3 c_s^2 H$ and 
$
 \partial^{ }_x\ln(s^{ }_\T a^3) = 0
$.

With this notation, the yield parameters for lepton minus baryon 
asymmetries evolve as 
\ba
 {Y}_a' - \frac{Y_\rmii{$B$}'}{3}
 & = & 
 \frac{4}{s^{ }_\T} 
   \int_{\vec{k}} \Bigl\{ 
   \bigl[f^{+}_{ } - \nF^{ }(\omega^{ }_1) \bigr]
   \, \widehat{B}^{+}_{(a)11} 
 + f^{-}_{ } \widehat{B}^{-}_{(a)11}
 - \nF^{ }(\omega^{ }_1) \bigl[1 - \nF^{ }(\omega^{ }_1)\bigr]\, 
  \widehat{A}^+_{(a)11}
 \Bigl\} 
 \nn 
 & + &   
 4 \, \tr \Bigl\{ 
   \bigl( Y^{+}_{\sH} - Y^+_\rmi{eq} \bigr)
    \langle \widehat{B}^{+}_{(a)\sH} \rangle^{ }_1 
 + Y^{-}_{\sH}\, \langle \widehat{B}^{-}_{(a)\sH} \rangle^{ }_1
 - \langle  \widehat{A}^+_{(a)\sH}  \rangle^{ }_2
 \Bigr\} \hspace*{4mm}
 \;, \la{evol1}
\ea
where $a \in \{e,\mu,\tau\}$ and $f^{\pm}_{ }$ are the distribution
functions of the keV-scale sterile neutrinos, for which no momentum
averaging is carried out, as their resonant production proceeds
one momentum mode at a time. 
The heavy components of the density matrix evolve as 
\ba
 (Y^{\pm}_\sH)' & = & 
   i \bigl[
   \langle  
    \diag \bigl( 
    \widehat{\omega}^{ }_2 ,
    \widehat{\omega}^{ }_3
    \bigr)
    - \widehat{H}^{+}_{\!\sH}
   \rangle^{ }_1\;,\; 
   Y^{\pm}_{\sH}
   \bigr]
 \; - \;  
   i \bigl[
     \langle \widehat{H}^{-}_{\!\sH} \rangle^{ }_1, 
     Y^{\mp}_{\sH}
   \bigr]
 \nn[2mm] 
 & + & 
 \bigl\{  
   \langle \widehat{D}^{\pm}_{\sH} \rangle^{ }_1 \,,\,
   Y^+_\rmi{eq} - Y^+_{\sH}
 \bigr\}
 \; - \;  
 \bigl\{  
   \langle \widehat{D}^{\mp}_{\sH} \rangle^{ }_1 \,,\,
   Y^-_{\sH}
 \bigr\}
 \; + \; 
 2\, \langle \widehat{C}^{\pm}_{\sH} \rangle^{ }_2 
 \;, \la{evol2}
\ea
whereas light components satisfy
\be
 ({\f}^{\pm})'  
  =  
 2 {D}^{\pm}_{11}\, \bigl[ \nF^{ }(\omega^{ }_1) - \f^{+}_{ } \bigr]
 - 
 2 {D}^{\mp}_{11}\, \f^{-}_{ }
 + 
 2 {C}^{\pm}_{11}\,
 \nF^{ }(\omega^{ }_1) \bigl[ 1 - \nF^{ }(\omega^{ }_1) \bigr] 
 \;. \la{evol3}
\ee
In \eq\nr{evol3}, $\omega^{ }_1 \equiv \sqrt{k^2(x) + M_1^2}$,
and $k(x)$ also appears in the coefficient functions.

Eqs.~\nr{evol1}--\nr{evol3} are 
parametrized by the rate coefficients
$\widehat{A}^{+}_{\I\J},...,\widehat{D}^\pm_{\I\J}$
and the effective Hamiltonians $\widehat{H}^{\pm}_{\sH}$. 
These are 
proportional to the second power of neutrino Yukawa couplings, 
and in some cases to chemical potentials
(e.g.\ $\widehat{A}^+_{\I\J}$, $\widehat{C}^\pm_{\I\J}$). 
Some important
coefficients are elaborated upon in \se\ref{se:lecs}, 
and all definitions and evaluations are explained
in ref.~\cite{simultaneous}.

%
\section{Low-energy constants}
\la{se:lecs}

The slow evolution equations in \eqs\nr{evol1}--\nr{evol3} contain a large
number of coefficients that capture the effects of the fast Standard
Model degrees of freedom. The physical values of the
coefficients are correlated, as all of them
originate from a certain retarded correlator of Standard Model 
operators~\cite{simultaneous,dbX}. Therefore the qualitative features 
of the solution only depend on the values of a few coefficients. 
Motivated by the analogy with effective field theories, 
we call these coefficients ``low-energy constants''.

For lepton chemical potentials large enough that all 
of dark matter gets produced, dark matter production peaks
at $T \sim 0.2$~GeV if $M^{ }_1 \sim 7$~keV~\cite{dmpheno}.
As most of the lepton asymmetries that were 
produced at higher temperatures 
get diluted away~\cite{simultaneous}, their production
should take place close to this temperature range~\cite{late}. 
Based on numerical integrations such as that described
in \se\ref{se:example}, we find that it is practical to fix 
$T = 0.5$~GeV for considering the values of the low-energy constants.

For producing large lepton asymmetries, CP-violation needs to
be resonantly enhanced. This means that the oscillations between
the GeV-scale species need to be slow, i.e.\ with a frequency
similar to the Hubble rate. The oscillation frequency 
is determined by three types of mass-squared differences: 
(i) Lagrangian parameters, $M^{2}_3 - M^{2}_2$; 
(ii) Higgs vev corrections $\sim \Delta\, h^2 v^2$; 
(iii) thermal corrections, which at low temperatures 
are $\propto G_\rmii{F}^2 T^4$, where
$G^{ }_\rmii{F}$ is the Fermi constant. 
At $T = 0.5$~GeV, corrections of types (i) and (ii) are the 
most important ones in absolute magnitude, however as it turns out that 
an exquisite cancellation is required between the different types of
corrections, the class~(iii) also plays a role 
(cf.\ \se\ref{ss:analytic}). 

Technically, slow oscillations require that, after subtracting
the trace part which has no effect, the Hamiltonian in 
the first commutator of \eq\nr{evol2} should
have a near-zero eigenvalue~\cite{late}. This 
gives us the first low-energy constant, which we define in 
the limit of thermal equilibrium and vanishing chemical potentials: 
\ba
 \langle \widehat{H}^{ }_\lambda \rangle^{ }_1 
  & \equiv & \lim_{\mu_i\to 0}
 \Bigl|\mathop{\mbox{eigenvalue}}\, 
  \Bigl(\langle 
          \diag \bigl( 
          \widehat{\omega}^{ }_2 ,
          \widehat{\omega}^{ }_3
          \bigr)
           - \widehat{H}^{+}_{\!\sH}
        \rangle^{ }_1
 -
   \frac{\mathbbm{1}}{2}\,
   \mbox{(trace)}
 \Bigr)
 \Bigr|^{ }_{\mathcal{J}\to 3 c_s^2 H^{ }_\T}
 \la{lec1} \\ 
 & = &
 \sqrt{ 
 \biggl[  
 \frac{
   \langle \omega^{ }_2 - \omega^{ }_3 \rangle^{ }_1
   - 
   \sum_a  ( |h^{ }_{2 a}|^2 -  |h^{ }_{3 a}|^2 )
   \langle U^{+}_{(a)\sH} \rangle^{ }_{1} 
 }{ 
   6 c_s^2 H^{ }_\T
 }  
 \biggr]^2 
 + 
 \biggl[  
 \frac{
   \sum_a  \re( h^{ }_{2 a} h^{*}_{3 a} )
   \langle U^{+}_{(a)\sH} \rangle^{ }_{1} 
 }{ 
   3 c_s^2 H^{ }_\T
 }
 \biggr]^2 
 }
 \;,  \nonumber
\ea
where $H^{ }_\T$ is the contribution to the Hubble rate from (thermal)
Standard Model degrees of freedom. 
The coefficient $U^+_{(a)\sH}$ captures the ``dispersive'' corrections
of types (ii) and (iii), and originates from the real part of a
matrix element of a retarded correlator~\cite{simultaneous}.  
 
It is not sufficient to have slow oscillations between the GeV-scale
sterile species, but suitable interaction rates need also to be present. 
Interaction rates should not be too large, otherwise the sterile
neutrinos stay close to equilibrium and furthermore an efficient 
washout of any lepton asymmetries takes place. But they should
not be too small either, 
otherwise no interesting dynamics takes place. In short,
interaction rates should be of the order of the Hubble rate. 
As a representative for a CP-even rate coefficient, we define  
\be
 \langle \widehat{\Gamma}^{ }_{\!\sH} \rangle^{ }_1
 \; \equiv \;
 \tr \langle \widehat{D}^{+}_\sH \rangle^{ }_1
 \Bigr|^{ }_{\mathcal{J}\to 3 c_s^2 H^{ }_\T}
 \; = \; 
 \frac{
   \sum_a \sum_{\I = 2,3} |h^{ }_{\I a}|^2 
   \langle Q^{+}_{(a)\sH} \rangle^{ }_{1} 
 }{ 
   3 c_s^2 H^{ }_\T
 }
 \;. \la{lec2}
\ee
The coefficient $Q^+_{(a)\sH}$ describes ``absorptive'' corrections, 
and originates from the imaginary part of a
matrix element of a retarded correlator~\cite{simultaneous}.  

Finally, the generation of lepton asymmetries necessitates the 
presence of CP-violating interaction rates. 
By inspecting solutions such as that
described in \se\ref{se:example}, we have found that a good 
representative for them is given by 
\be
 \langle \widehat{\Gamma}^{ }_\rmi{osc} \rangle^{ }_1
 \; \equiv \; 
 \lim_{\mu_i\to 0} \langle i \widehat{D}^{+}_{23} \rangle^{ }_1
 \Bigr|^{ }_{\mathcal{J}\to 3 c_s^2 H^{ }_\T}
 \; = \; 
 \frac{
   \sum_a \im( h^{ }_{2 a} h^{*}_{3 a} )  
   \langle Q^{-}_{(a)\sH} \rangle^{ }_{1} 
 }{ 
   3 c_s^2 H^{ }_\T
 }
 \;. \la{lec3}
\ee
Here $Q^-_{(a)\sH}$ describes a helicity-antisymmetrized
absorptive correction. Again, \eq\nr{lec3} should be
of order unity for an efficient production
of lepton asymmetries.   

%
\section{Parameter scans}
\la{se:scans}

The purpose of this section is to determine the possible values 
of the low-energy constants in \eqs\nr{lec1}--\nr{lec3}. We start
with a numerical scan in \se\ref{ss:numerical}, and subsequently
explain the results  
through analytic estimates in \se\ref{ss:analytic}. 

%
\subsection{Numerical results}
\la{ss:numerical} 

In the following we employ the Casas-Ibarra parametrization~\cite{ci}, 
which ensures that active neutrino mass differences and mixing angles
take their observed values. As only two mass differences are known, 
we can account for the observed values through the two GeV-scale sterile
neutrinos. With 
\be
 M^{ }_\rmii{M} \equiv \left( 
 \begin{array}{cc}
  M^{ }_2 & 0 \\ 
  0 & M^{ }_3
 \end{array}
 \right)
 \;, \quad
 R \equiv 
 \left( 
 \begin{array}{cc}
  \;\, \cos z & \sin z \\ 
  - \sin z & \cos z
 \end{array}
 \right)
 \;, \quad z \in \mathbbm{C}
 \;, \la{ci1}
\ee
this is obtained through the relation
\be
 h^{ } = - i 
 \sqrt{M^{ }_\rmii{M}}\, 
  R^{ }_{ }\, 
  P^{ }_{ }\,
 \underbrace{ \sqrt{m^{ }_\nu}\, 
  V^{\dagger}_\rmii{PMNS} }_{\rmi{data}}\, 
  \frac{\sqrt{2}}{v}
  \;, \la{ci2}
\ee
where $m^{ }_\nu$ is a diagonal matrix containing the active
neutrino masses, $V^{ }_\rmii{PMNS}$ is the Dirac-like
mixing matrix
(which contains the complex phase $\delta$),  
$v\simeq 246$~GeV is the Higgs vev, and 
\be
 P^{ }_\rmii{\,NH} \equiv 
 \left( 
 \begin{array}{ccc}
   0 & e^\rmi{$-i \phi_1$} & 0 \\ 
   0 & 0 & 1 
 \end{array}
 \right)
 \;, \quad
 P^{ }_\rmii{IH} \equiv
 \left( 
 \begin{array}{ccc}
   1 & 0 & 0 \\ 
   0 & e^\rmi{$-i \phi_1$} & 0 
 \end{array}
 \right)
 \;. \la{ci3}
\ee
Here NH and IH stand for the normal and inverted hierarchy, respectively. 

\begin{figure}[t]

\hspace*{-0.1cm}
\centerline{%
 \epsfysize=7.5cm\epsfbox{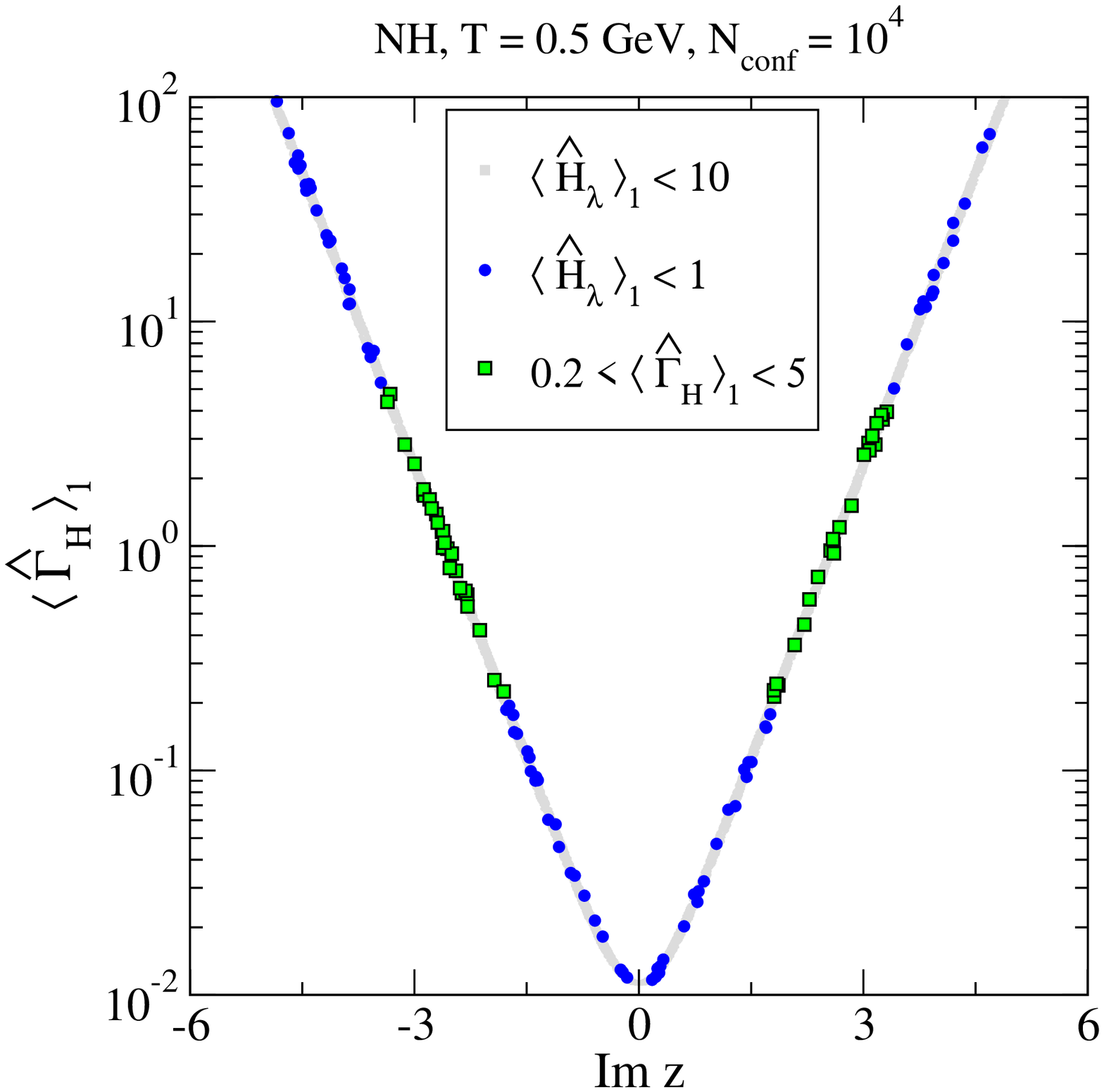}%
 \hspace{0.5cm}%
 \epsfysize=7.5cm\epsfbox{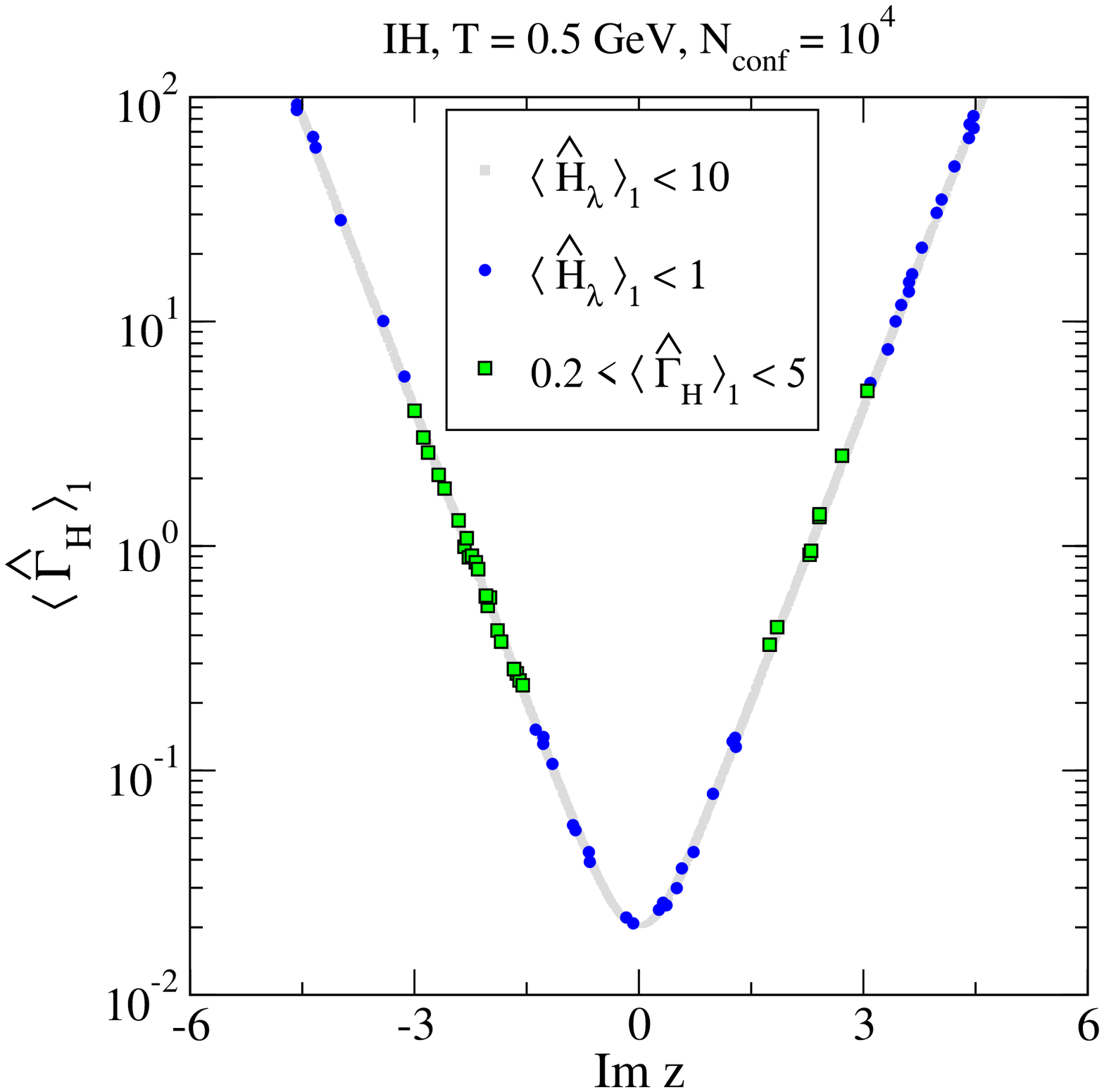}%
}

\caption[a]{\small
 The distribution of
 $\langle \widehat{\Gamma}^{ }_{\!\sH} \rangle^{ }_1$
 as a function of $\im z$,  
 with increasingly stringent constraints as indicated by the legend
 (left: normal hierarchy, right: inverted hierarchy).
 $\langle \widehat{\Gamma}^{ }_{\!\sH} \rangle^{ }_1$
 is strongly correlated with $\im z$, with the influence
 of the other parameters corresponding to the line width.  
 For a rate not too slow or fast, we restrict to 
 $0.2 < \langle \widehat{\Gamma}^{ }_{\!\sH} \rangle^{ }_1 < 5.0$
 (the points indicated with the green squares).
 As visible by the density of points, valid parameters 
 are somewhat rarer for the inverted hierarchy.  
}

\la{fig:scans_1}
\end{figure}

To proceed we choose a representative value 
$ 
M^{ }_{\sH} = 2$~GeV, and carry out  
a scan in the five-dimensional parameter space spanned
by $\Delta M \equiv M^{ }_3 - M^{ }_2$, 
$\re z$, $\im z$, $\delta$, and $\phi^{ }_1$.\footnote{%
 For $\re z$ we restrict to a region continuously connected
 to $\pi/2$ (another lies at around $\re z = -\pi/2$).
 } 
A random sample of data points is generated,  
with a logarithmic distribution in $\Delta M$ and 
a flat one in the other parameters (restricting to $|\im z|\le 6.0$). 
The distributions 
in \figs\ref{fig:scans_1}--\ref{fig:scans_3} are based on $\sim 10^4$
``accepted'' points that satisfy the weak criterion 
$
 \langle \widehat{H}^{ }_\lambda \rangle^{ }_1  < 10 
$
in terms of the low-energy constant defined in \eq\nr{lec1}. 
The criteria for a successful dark matter scenario are stronger than 
this, and reduce the data set to $\sim 10^2$ points. We should
clarify that these particular values are chosen for ease of illustration;
many more points can be generated with minor cost but do not 
change the conclusions.  

First, let us constrain $\im z$ by considering
$
  \langle \widehat{\Gamma}^{ }_{\!\sH} \rangle^{ }_1
$
from \eq\nr{lec2}. As shown in \fig\ref{fig:scans_1}, there is 
a near-perfect correlation of 
$
  \langle \widehat{\Gamma}^{ }_{\!\sH} \rangle^{ }_1
$
and
$\im z$.
We restrict to 
$0.2 < \langle \widehat{\Gamma}^{ }_{\!\sH} \rangle^{ }_1 < 5.0$, 
which delineates the range allowed for $|\im z|$ accordingly.

\begin{figure}[t]

\hspace*{-0.1cm}
\centerline{%
 \epsfysize=7.5cm\epsfbox{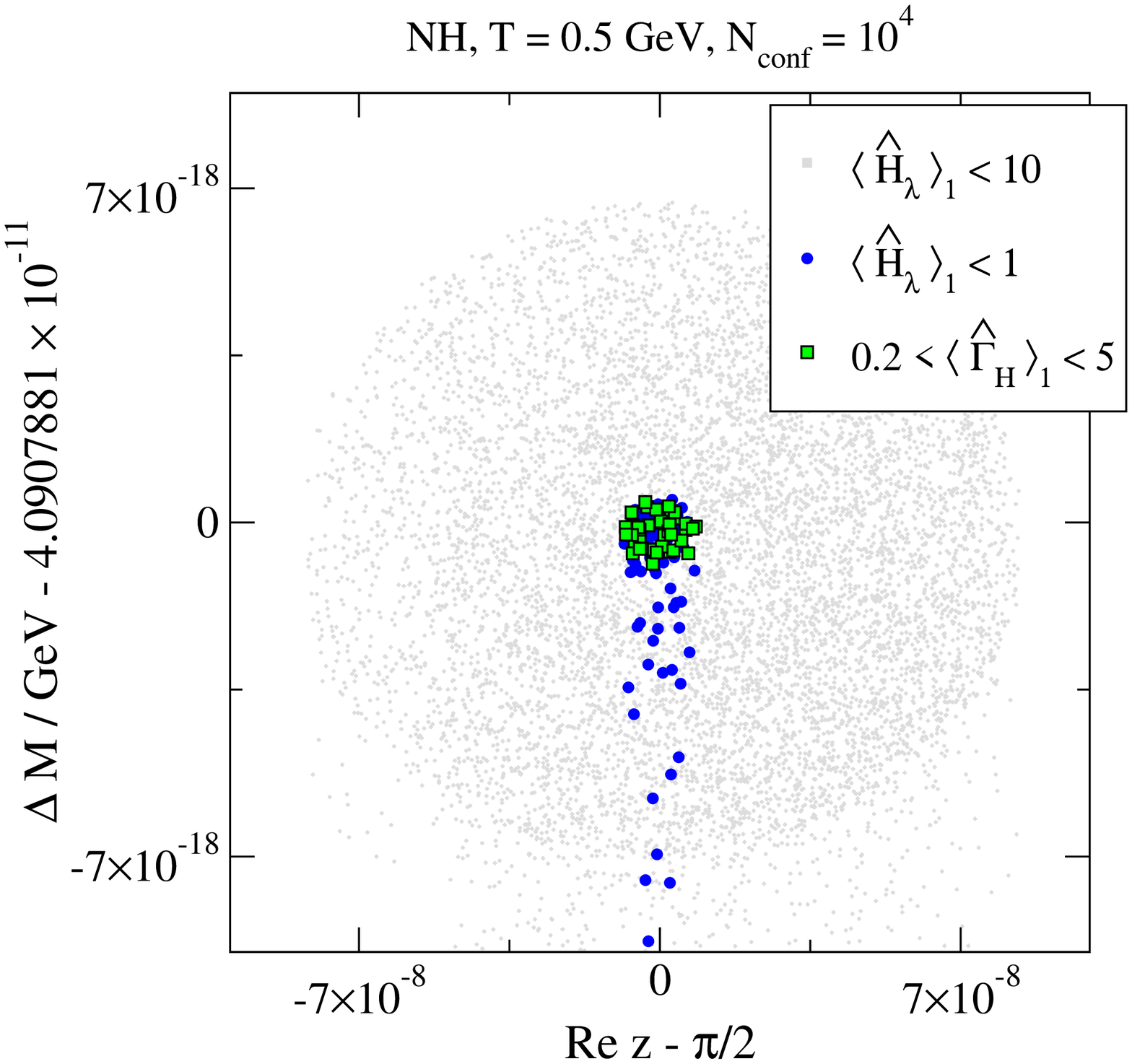}%
 \hspace{0.5cm}%
 \epsfysize=7.5cm\epsfbox{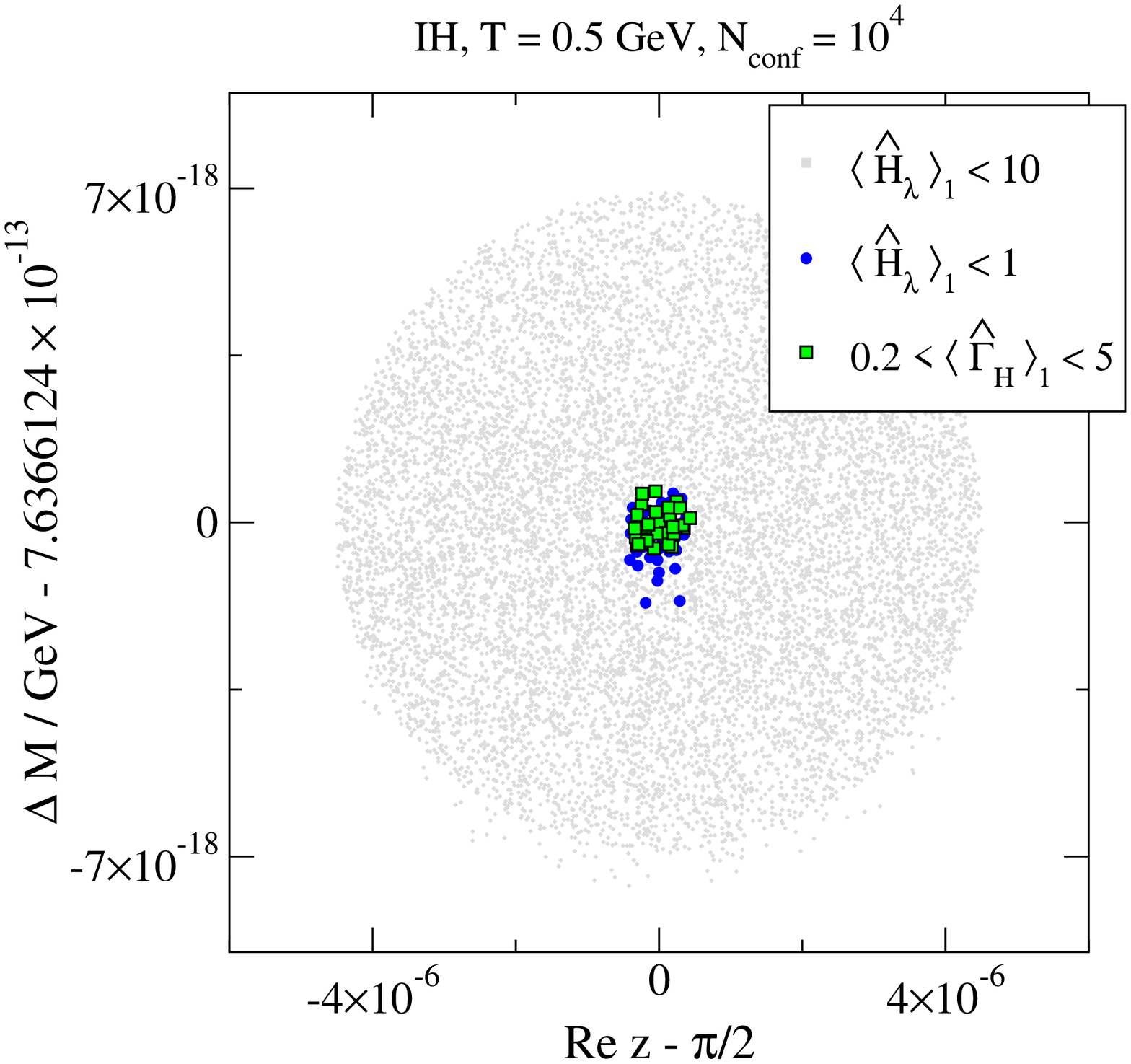}%
}

\caption[a]{\small
 Points satisfying the increasingly stringent constraints indicated 
 by the legend (cf.\ \se\ref{ss:numerical}), 
 in the plane of 
 $\re z$ and $\Delta M$
 (left: normal hierarchy, right: inverted hierarchy). 
 The narrow axis ranges illustrate
 the extraordinary degree of fine-tuning
 that is needed for realizing the desired scenario.
}

\la{fig:scans_2}
\end{figure}

Second, we constrain
$
 \langle \widehat{H}^{ }_\lambda \rangle^{ }_1   
$
from \eq\nr{lec1} to lie within the range 
$
 \langle \widehat{H}^{ }_\lambda \rangle^{ }_1  < 1 
$.
A full solution such as the one in \se\ref{se:example} shows
that this is necessary for obtaining sufficient resonant 
enhancement of lepton asymmetry generation. 
As shown in \fig\ref{fig:scans_2}, 
this cuts the range allowed for 
$\Delta M$ and $\re z$ to an extremely narrow domain, 
which becomes an ellipse once the permitted range of 
$
  \langle \widehat{\Gamma}^{ }_{\!\sH} \rangle^{ }_1
$
is also taken into consideration.

\begin{figure}[t]

\hspace*{-0.1cm}
\centerline{%
 \epsfysize=7.5cm\epsfbox{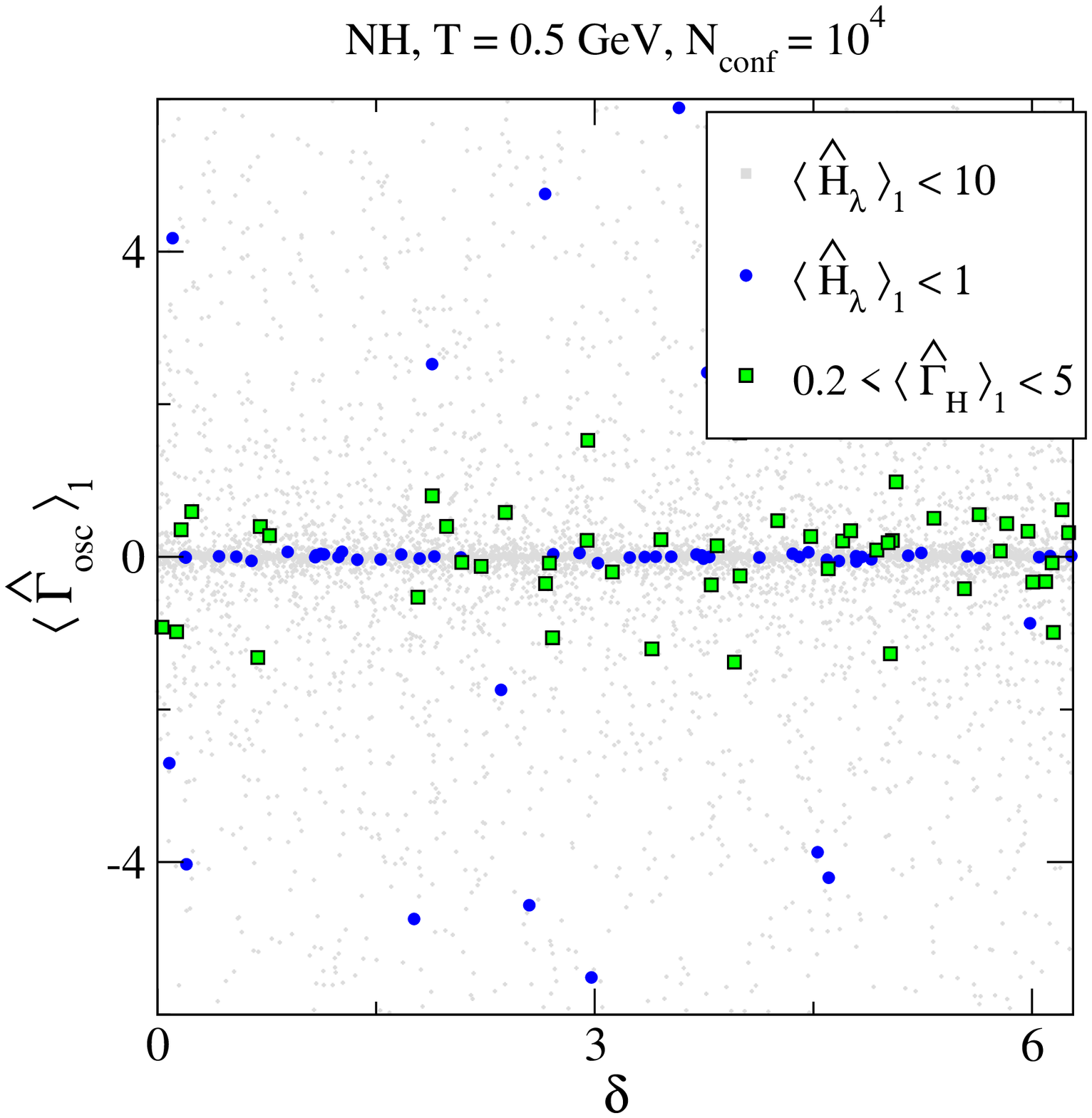}%
 \hspace{0.5cm}%
 \epsfysize=7.5cm\epsfbox{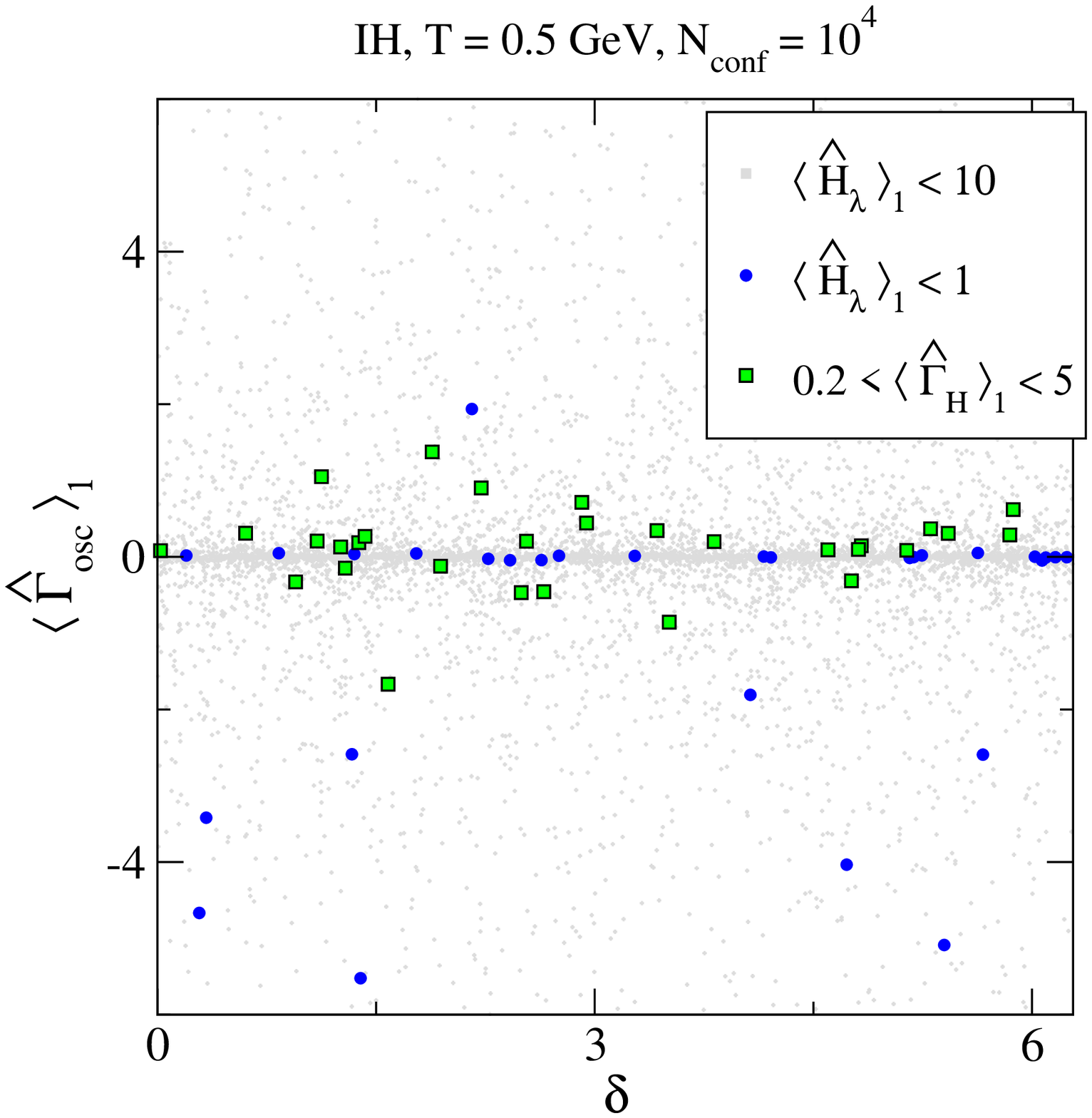}%
}

\caption[a]{\small
 The distribution of
 $\langle \widehat{\Gamma}^{ }_\rmi{osc} \rangle^{ }_1$
 as a function of $\delta$
 (left: normal hierarchy, right: inverted hierarchy).
 The effectively flat shape implies that dark matter production 
 puts no constraint on $\delta$, and that therefore $\delta$ can be tuned
 to obtain the correct baryon asymmetry. 
 The same is true for $\phi^{ }_1$.  
}

\la{fig:scans_3}
\end{figure}

Finally, in \fig\ref{fig:scans_3}, we show the low-energy constant
$
  \langle \widehat{\Gamma}^{ }_\rmi{osc} \rangle^{ }_1
$
from \eq\nr{lec3} as a function of the CP-violating phase $\delta$.
The plot looks identical for $\phi^{ }_1$. Practically
no correlation is observed, implying that many values of 
$\delta$ and $\phi^{ }_1$ are available. We have checked that 
these can then
be tuned to produce the correct baryon asymmetry at $T \sim 130$~GeV.  

%
\subsection{Analytic estimates}
\la{ss:analytic}

The purpose of this section is to show that 
the features observed in \figs\ref{fig:scans_1}--\ref{fig:scans_3}
can be understood analytically. To this end, let us start 
by noting that eqs.~\eqref{lec1}--\eqref{lec3} involve summations over
the active flavour $a$ of a combination of Yukawa couplings and the rates
$\langle U^{+}_{(a)\sH} \rangle^{ }_{1}$ 
and $\langle Q^{\pm}_{(a)\sH} \rangle^{ }_{1} $.
At vanishing chemical potentials, only the masses of 
the charged leptons can lead to $a$-dependence. 
In the temperature range that we are interested in, this effect
is in principle substantial for the $\tau$-lepton, however it 
originates through thermal corrections, which are suppressed. 
Therefore it is a good first approximation to treat
the rates as flavour-independent. 
Then the active flavour sum over the neutrino Yukawas takes the form
\begin{align}
  \label{yukawa}
  & \sum_a h^{ }_{\I a} h^{*}_{\J a} = \\
  & \nonumber \frac{1}{v^2}
 \Biggl\{ 
 \hspace{-2mm}\begin{array}{cc}
  M^{ }_2
 \left[-\delta m \cos(2 \re z)+ \bar m\cosh(2\im z) \right]
  &\hspace{-4mm}
 \sqrt{ M^{ }_2 M^{ }_3 }
 \left[ \delta m \sin(2\re  z)+i  \bar m \sinh(2\im z) \right]
  \\
  \sqrt{ M^{ }_2 M^{ }_3}
 \left[ \delta m \sin(2\re  z)-i  \bar m \sinh(2\im z) \right]
  &\hspace{-2mm}
  M^{ }_3 
 \left[ \delta m \cos(2 \re z)+ \bar m\cosh(2\im z) \right]
 \end{array}\hspace{-2mm} \Biggr\}^{ }_{\!\I\J}
 \!\! ,
\end{align}
where 
\begin{align}
   \bar m \;\equiv\;
   &
   \sqrt{\Delta m_{31}^2}+\sqrt{\Delta m_{21}^2} \;, 
   &
   \delta m \;\equiv\;
   &
   \sqrt{\Delta m_{31}^2}-\sqrt{\Delta m_{21}^2} 
   &
   \mathrm{(NH)} 
   \;,\\
   \bar m \;\equiv\;
   &
   \sqrt{\Delta m_{23}^2}+\sqrt{\Delta m_{23}^2-\Delta m_{21}^2} \;,  
   &
   \delta m \;\equiv\;
   &
   \sqrt{\Delta m_{23}^2}-\sqrt{\Delta m_{23}^2-\Delta m_{21}^2} 
   &
   \mathrm{(IH)}
   \;.
\end{align}
Here $\Delta m^2_{ij} \equiv m_i^2 - m_j^2$ are active neutrino mass
differences, and we have employed the combinations 
appearing in standard fits 
($
 \Delta m^2_{21} = \Delta m^2_\rmi{sol}
$, 
$
 \Delta m^2_{31} |^{ }_\rmii{NH} \approx
 \Delta m^2_{23} |^{ }_\rmii{IH} = \Delta m^2_\rmi{atm}
$). 

Let us now take eq.~\eqref{lec1} 
and study its vacuum contributions, i.e.\ those of types (i) and (ii)
in the language of sec.~\ref{se:lecs}. This corresponds to taking
\begin{equation}
  \langle U^{+}_{(a)\sH} \rangle^{ }_{1} 
  \to 
  -\frac{v^2}{2}\, \langle \omega_{\sH}^{-1} \rangle^{ }_{1}
  \;. 
   \label{zeroTU}
\end{equation}
We can also write  
\begin{equation}
  \langle \omega^{ }_2 - \omega^{ }_3 \rangle^{ }_1 =
  \left\langle 
  -\frac{2 \Delta M\, M^{ }_{\sH}}
        {\omega^{ }_2 + \omega^{ }_3}\right\rangle^{ }_1
   \approx 
  -\Delta M\, M^{ }_{\sH}\langle \omega_{\sH}^{-1} \rangle^{ }_{1}
  \;.
  \label{deltaom}
\end{equation}
Plugging eqs.~\eqref{zeroTU} and \eqref{deltaom} into \eq\nr{lec1} 
and using \eq\nr{yukawa}, the argument of the square root is 
a second order polynomial in $\Delta M$. 
It has a minimum for $\Delta M= {\Delta M}^{ }_\rmi{min} $, where
\begin{eqnarray}
  \label{minimum}
   {\Delta M}^{ }_\rmi{min} &=&
   \frac{-2 \delta m\, M^{ }_{\sH} 
           \cos (2 \re z) 
          \left[ 2 M^{ }_{\sH}+\bar m \cosh (2 \im z) \right] }
  {\left[ 2 M^{ }_{\sH}+\bar m \cosh (2 \im z)\right]^2
  -\delta m^2 \sin^2 (2  \re z)}\\
  &\approx&
  -\delta m \cos (2 \re z)
  \, \biggl[ 
    1-\frac{\bar m \cosh(2\im z)}{2M_{\sH}}
     +\mathcal{O}\left(\frac{m_\nu^2}{M^2_{\sH}} \right)
  \biggr] \;.
  \label{minimumexpand}
\end{eqnarray}
In the final approximation 
we have kept the first correction in $m^{ }_\nu/M^{ }_{\sH}$,
because it can be enhanced by 
the possibly large hyperbolic function.\footnote{%
 We note that 
 for very large Yukawas the Casas-Ibarra parametrization
 needs to be generalized~\cite{donini}. 
 }

Evaluating \eq\nr{lec1} at the minimum given by \eq\nr{minimumexpand}, 
we obtain
\begin{equation}
  \langle \widehat{H}^{ }_\lambda \rangle^{ }_1 
   \big|_{\Delta M = \Delta M_\rmii{min} }^{\mathrm{(i)+(ii)}}
   \; \approx \;  
   \frac{\delta m\,M_{\sH}|\sin(2\re z)|
   \langle \omega_{\sH}^{-1} \rangle^{ }_{1}
   }{6 c_s^2 H^{ }_\T}
  \, 
  \biggl[ 
   1 + \mathcal{O}\biggl( \frac{ m_\nu^2}{ M^2_{\sH} } \biggr)
  \biggr]
  \;, 
  \label{minlambda}
\end{equation}
where we have again dropped higher-order terms in $ m_\nu/ M_{\sH}$.
Noting that the values $\re z = 0$~mod~$\pi$ 
are excluded by requiring $\Delta M$
to be positive in \eq\nr{minimumexpand}, we find that, 
as first shown in ref.~\cite{late}, the cancellation 
between the (i) and (ii) contributions is largest ---
and the vacuum part of $\langle \widehat{H}^{ }_\lambda \rangle^{ }_1$ 
is smallest --- for $\re z=\pm\pi/2$. 

To see the extent of the cancellation, we may note that
the (i) contribution evaluates for $\Delta M=\Delta M^{ }_\rmi{min}$ to 
\begin{equation}
  \langle \widehat{H}^{ }_\lambda \rangle^{ }_1 
   \big|_{\Delta M = \Delta M_\rmii{min} }^{\mathrm{(i)}}
   \; \approx \;  
   \frac{\delta m\,M_{\sH}|\cos(2\re z)|
   \langle \omega_{\sH}^{-1} \rangle^{ }_{1}
   }{6 c_s^2 H^{ }_\T}
  \, 
  \biggl[ 
   1 + \mathcal{O}\biggl( \frac{ m_\nu^2}{ M^2_{\sH} } \biggr)
  \biggr]
  \;. 
  \label{minlambdai}
\end{equation}
As $\re z$ approaches $\pm\pi/2$,
the difference between \eqs\nr{minlambda} and \nr{minlambdai}
is maximized. It is for this reason that thermal corrections to 
$  \langle \widehat{H}^{ }_\lambda \rangle^{ }_1  $, of type~(iii)
in the language of \se\ref{se:lecs} and  
ordinarily suppressed by $\sim (T/v)^4$,
are relatively speaking 
important when $\Delta M \approx \Delta M^{ }_\rmi{min}$.

For fixed $\re z$, the value of \eq\nr{minlambda} 
increases with decreasing temperature,
because the Hubble rate shrinks and 
$M^{ }_\sH\langle \omega_{\sH}^{-1} \rangle^{ }_{1}$
depends mildly on $T$.
The thermal contribution~(iii), on the other hand, 
increases with temperature. Hence,
$\langle \widehat{H}^{ }_\lambda \rangle^{ }_1$  exhibits 
a minimum as a function of temperature. The condition
$\langle \widehat{H}^{ }_\lambda \rangle^{ }_1<1$ at $T=0.5$ GeV 
ensures that this minimum is located 
where dark matter production is most efficient.

Fig.~\ref{fig:scans_2} can now be understood as follows. 
For $|\im z|\lsim 3$, the $\cosh(2\im z)$ correction 
in eq.~\eqref{minimumexpand} is negligible, so that 
${\Delta M}^{ }_\rmi{min} \approx -\delta m \cos (2 \re z) \approx \delta m$. 
This explains why most of the accepted points are symmetrically
distributed around $\re  z=\pi/2$, $\Delta M=\delta m$. 
The outliers below the center correspond to 
$|\im z| > 3$, for which 
$\langle \widehat{\Gamma}^{ }_{\!\sH}\rangle^{ }_1\gsim 5$.
As shown by eq.~\eqref{minimumexpand}, the correction 
from $\cosh(2\im z)$ indeed reduces $\Delta M^{ }_\rmi{min}$.

As far as the other parameters go, eq.~\eqref{yukawa}
shows that $ \langle \widehat{\Gamma}^{ }_{\!\sH}\rangle^{ }_1$ and
$ \langle \widehat{\Gamma}^{ }_\rmi{osc} \rangle^{ }_1$ are  
proportional to $(\bar m M^{ }_{\!\sH}/v^2) \cosh(2\im z)$
and $(\bar m M^{ }_{\!\sH}/v^2) \sinh(2\im z)$, respectively. 
Indeed, fig.~\ref{fig:scans_1} is effectively
a plot of the $\cosh(2\im z)$ dependence, with the width 
of the line an indicator of the goodness of the 
flavour-independence approximation. 
The same approximation makes
$ \langle \widehat{\Gamma}^{ }_\rmi{osc} \rangle^{ }_1$ 
independent of the phases $\delta$ and $\phi^{ }_1$, 
as illustrated in fig.~\ref{fig:scans_3}.

We end by noting that many more digits are shown on the vertical
axes in \fig\ref{fig:scans_2} than there are significant
digits in $\delta m$, or that correspond to 
the theoretical uncertainties of the 
computation. Therefore the location of the optimal value of 
$\Delta M$ is subject to uncertainty, however the degree of 
fine-tuning around this optimal value should be less so. 

%
\section{Example of a successful solution}
\la{se:example}

We end by offering a ``proof of existence'' for a successful dark 
matter scenario. For this we consider the normal hierarchy of neutrino
masses, and choose the parameters entering \eq\nr{ci1} to lie within
the domain found in \fig\ref{fig:scans_2}(left), {\it viz.}  
$
 M^{ }_{\sH} = 2
$~GeV,  
$
 \Delta M = 4.0907881 \times 10^{-11}~\mbox{GeV} 
$, 
$
 z = 1.570796327  + 3.0i 
$, 
$
 \delta = - 1.88496 
$, 
$ 
 \phi^{ }_1 = - 0.07037 
$.
The low-energy constants of \eqs\nr{lec1}--\nr{lec3} evaluate at
$T = 0.5$~GeV to 
\be
 \langle \widehat{H}^{ }_\lambda \rangle^{ }_1 = 0.397
 \;, \quad
 \langle \widehat{\Gamma}^{ }_{\!\sH} \rangle^{ }_1 = 2.323
 \;, \quad
 \langle \widehat{\Gamma}^{ }_\rmi{osc} \rangle^{ }_1 = - 0.800
 \;.
\ee
We have checked that these parameters leave behind the correct baryon
asymmetry after its freeze-out 
at $T \approx 130$~GeV~\cite{sphaleron},\footnote{%
  In fact, the value is $\YB^{ } = 1.65 \times 10^{-10}$, 
  i.e.\ larger than the observed $\YB^{ } = 0.87 \times 10^{-10}$, but the 
  result is diluted by $\sim 10$\% due to the entropy release resulting
  from the freeze-out and decays of the GeV-scale sterile neutrinos. 
  The remainder could be adjusted by tuning the CP-violating phases
  $\delta$ and $\phi^{ }_1$.}
and concentrate in the following on dark matter production. 
As dark matter production takes place at temperatures close to the QCD
crossover and therefore contains substantial hadronic uncertainties, 
we choose to be conservative and  
even overproduce dark matter moderately. 

As initial values for the lepton asymmetries 
we insert a ``fixed-point'' solution obtained in
accordance with ref.~\cite{degenerate}, 
$Y^{ }_a - \YB^{ }/3 \approx 2.0 \times 10^{-8}$
$\forall a$ at $T \approx 5.6$~GeV.
The dark matter mass is set to $M^{ }_1 = 7$~keV 
and its Yukawa couplings to 
$|h^{ }_{1a}| = 1.6 \times 10^{-13}$ $\forall a$.

\begin{figure}[t]

\hspace*{-0.1cm}
\centerline{%
 \epsfysize=5.0cm\epsfbox{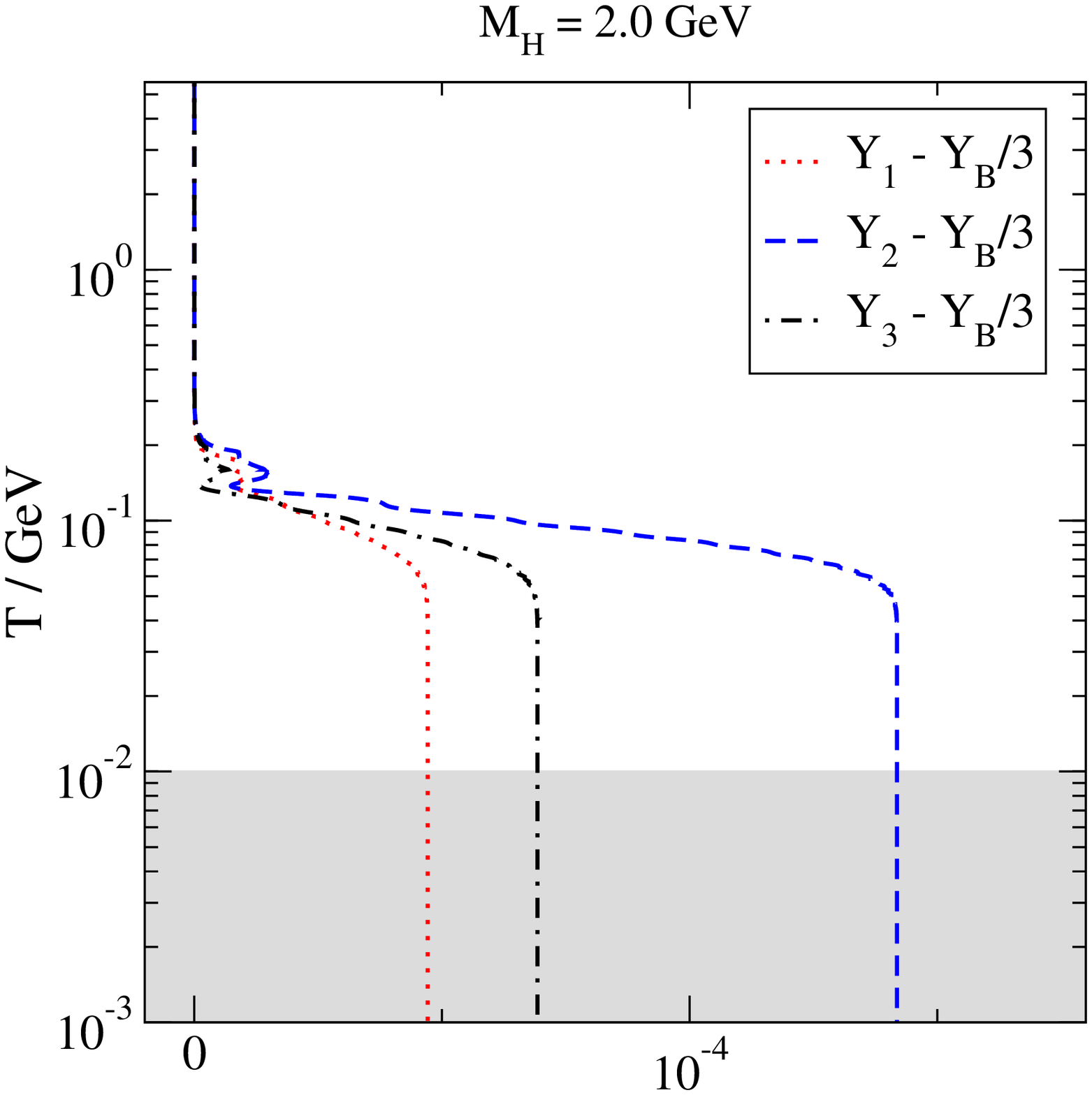}%
 \hspace{0.5cm}%
 \epsfysize=5.0cm\epsfbox{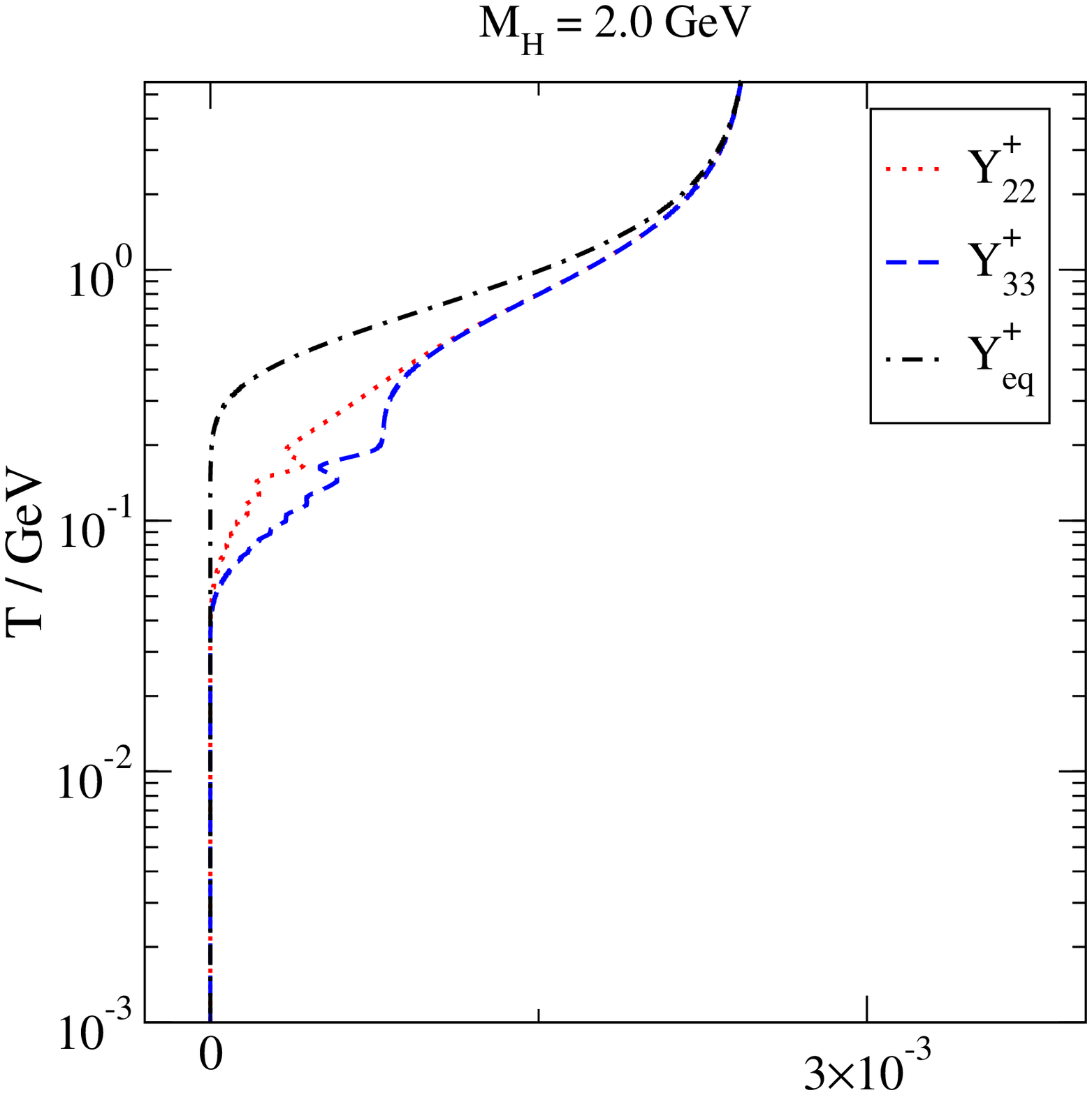}
 \hspace{0.5cm}%
 \epsfysize=5.0cm\epsfbox{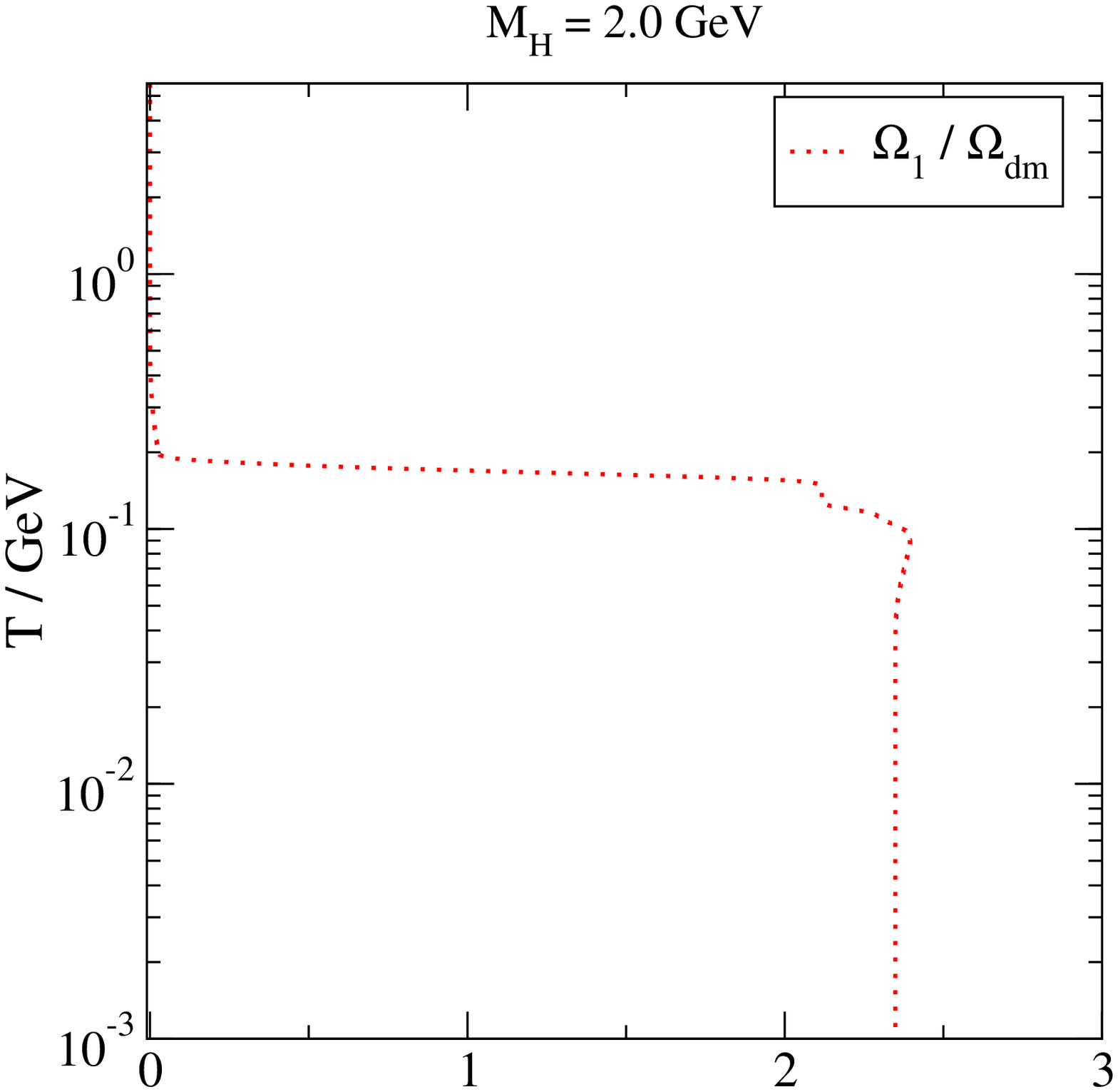}
}

\caption[a]{\small
 Example of a solution leading to a substantial dark matter abundance. 
 Left: 
 the evolution of lepton asymmetries. The grey band indicates
 the region in which active neutrino oscillations, not 
 included in our solution, 
 are expected to lead to flavour equilibration~\cite{eq1,eq2}. 
 Middle: 
 diagonal components of the GeV-scale density matrix, 
 compared with the equilibrium value  $Y^+_\rmii{eq}$
 from \eq\nr{ansatz}. 
 Right: 
 the fraction of dark matter that keV-scale sterile
 neutrinos account for, obtained from \eq\nref{fracOmegadm}. 
 The minor decrease after obtaining the peak value
 is due to entropy dilution. 
}

\la{fig:Y_Hpx}
\end{figure}

The evolutions of the lepton asymmetries 
are shown in \fig\ref{fig:Y_Hpx}(left), whereas 
the diagonal components of the GeV-scale density matrix 
can be found in \fig\ref{fig:Y_Hpx}(middle).
Lepton asymmetries of absolute magnitude $\sim 10^{-4}$ can 
indeed be obtained, thanks to the fact that the density
matrix falls out of equilibrium in a region where 
$
 \langle \widehat{\Gamma}^{ }_\rmi{osc} \rangle^{ }_1 
$ is substantial. 

When the temperature falls much below the mass of 
the GeV-scale sterile neutrinos, their density 
becomes Boltzmann-suppressed. Therefore the lepton 
number washout rate switches off, and lepton asymmetries become
constant. At temperatures below $T \sim 10$~MeV, 
lepton asymmetries would however evolve again, as  
active neutrino oscillations are expected to enforce flavour
equilibrium~\cite{eq1,eq2}. This effect has not been  
accounted for here, so the corresponding domain has been shaded 
in \fig\ref{fig:Y_Hpx}(left). However, this does not
affect dark matter production which has ceased at $T \sim 100$~MeV.  

The dark matter abundance is shown in \fig\ref{fig:Y_Hpx}(right).
{}From the helicity-averaged distribution $f^{+}_{ }$ we compute 
the yield $Y^{+}_{11} \equiv \int_{\vec{k}^{ }_{ }} f^+_{ } / s^{ }_{\T}$
and subsequently estimate
\be
 \frac{ \Omega^{ }_{1} }{\Omega^{ }_\rmi{dm}}
 \approx 
 4.57\times Y^{+}_{11} \times \frac{M^{ }_1}{\mbox{eV}} 
 \;. \la{fracOmegadm}
\ee
The data overshoot the correct value modestly, an effect which
can easily be removed by a minuscule change of parameters such
as $\Delta M$ or $\re z$ (cf.\ \fig\ref{fig:scans_2}).

%
\section{Conclusions}
\la{se:concl}

By solving a set of coupled evolution equations for sterile neutrino
density matrices and Standard Model lepton asymmetries at temperatures
between 5.6~GeV and 1~MeV, we have scrutinized a proposal made in 
ref.~\cite{late} that this dynamics could lead to the generation of 
the correct dark matter abundance. As shown in \fig\ref{fig:Y_Hpx}, we 
are happy to confirm the idea. 

General features of solutions of the evolution equations
can be understood in terms of a small
number of low-energy constants, defined in \se\ref{se:lecs}. As discussed 
in \se\ref{ss:analytic} and illustrated numerically in \fig\ref{fig:scans_2}, 
it requires an exquisite degree of fine-tuning to set the low-energy 
constants in a domain leading to a solution like in \fig\ref{fig:Y_Hpx}. 
This fine-tuning concerns particularly the Lagrangian mass difference
$\Delta M = M^{ }_3 - M^{ }_2$, whose value needs to be small 
($< 0.1$~eV) and precisely chosen (to within $\sim$~ppm), 
as well as the Casas-Ibarra angle $\re z$, which needs to be close
to $\pm \frac{\pi}{2}$, and tuned to within a similar relative precision. 
We note that after a cancellation of $\Delta M$ against Higgs vev and 
thermal corrections to the required degree, the physical mass splitting
is $< 10^{-7}$~eV. 

In case these fine-tunings are not present, 
the dynamics in the chosen mass range generically leads to the generation 
of $\sim 5 - 10\%$ of the observed dark matter abundance~\cite{simultaneous}. 

We should end with a word of warning, which 
is simultaneously a call for further work. The low-energy constant 
$ \langle \widehat{H}^{ }_\lambda \rangle^{ }_1 $
defined in \eq\nr{lec1} is small only if major cancellations
between different contributions take place
(cf.\ \se\ref{ss:analytic}). However, the individual
contributions are only known within leading-order accuracy in 
Standard Model couplings. Higher-order corrections, even if small
for each contribution separately, are expected to be much larger
than the remainder. Therefore, for {\em fixed}
input parameters, the cancellation may be lifted by higher-order
corrections. However, it should still be reachable if the 
input parameters are tuned slightly. 
It would be interesting to confirm this expectation by determining
next-to-leading order corrections to thermal masses
(which we have parametrized through 
the coefficient $U^{+}_{(a)\sH}$), 
an exercise that has so far been carried out only for the 
helicity-conserving part of the interaction rates
$Q^{\pm}_{(a)\sH}$ at temperatures
somewhat below the electroweak crossover~\cite{nlo_width}.

%
\section*{Acknowledgements}

We are grateful to M.~Shaposhnikov 
and I.~Timiryasov for helpful comments on the manuscript.
M.L.\ was partly supported by the Swiss National Science Foundation
(SNF) under grant 200020B-188712.

%
\appendix
\renewcommand{\thesection}{\Alph{section}} 
\renewcommand{\thesubsection}{\Alph{section}.\arabic{subsection}}
\renewcommand{\theequation}{\Alph{section}.\arabic{equation}}

%

\small{
 
}

\end{document}